\documentclass[prb,showpacs,twocolumn]{revtex4}

\usepackage{graphicx,latexsym}

\begin{document}
\title{Magnon Modes for Thin Circular Vortex State Magnetic Dot}
\author{B.A.Ivanov}
\email{bivanov@i.com.ua}
\affiliation{Institute of Magnetism,
NASU, 36(B) Vernadskii avenue, Kiev 03142, Ukraine}
\author{C.E.Zaspel}
\affiliation{Department of Physics, Montana State
University/Bozeman, Bozeman, MT 59717, USA }
\date{\today }

\begin{abstract}
The magnetization in a magnetic microdot made from soft magnetic
materials can have a vortex-like ground state structure resulting
from competition between the exchange and dipolar interactions.
Normal mode magnon frequencies for such dots are calculated taking
into account both exchange and magnetostatic effects. The presence
of a low-lying mode as well as doublet structure with small
splitting is demonstrated. Estimates of the mode frequencies for
permalloy dots are obtained, and the possibility of experimental
detection of such modes is discussed.

\end{abstract}

\pacs{75.60.Jp, 75.10.Hk, 74.40. Cx }

\maketitle

Nanostructures fabricated from soft magnetic materials are of
interest both in basic and applied magnetics with potential
applications including high-density magnetic storage using an
array of dot structures.\cite{Smyth1, Maeda2} In the following we
will consider the dynamics of thin magnetic dots, with the
important parameter in relation to the dot radius, $R$ and the dot
thickness,$L$ being the exchange length, $l_0=\sqrt{A/ 4\pi
M_s^2}$ , where $A$ is the inhomogeneous exchange constant and
$M_s$ is the saturation magnetization. It has been shown
\cite{Miramond3} that under certain conditions a vortex structure
will be stable because of competition between the exchange and
dipole interactions, and it is expected that these non-uniform
states will drastically change the dynamic and static properties
of a dot. For this reason dynamic properties for thin ($L \ll R$)
magnetic dots in the vortex state are studied by determination of
the spin wave normal mode frequencies. The method used is based on
the determination of the of vortex-magnon scattering
amplitude,\cite{Ivanov4} where the non-local dipolar interaction
is included by calculation of the magnetostatic energy arising
from spin waves on the vortex background. A calculation of the
discrete spin-wave spectrum leads to modes labeled by an azimuthal
number, $m$, and a principal number, $n$. The magnon modes for the
dot are compared with modes in the uniformly magnetized
dot.\cite{Guslienko5} For vortex state dots a quasi-Goldstone mode
($m=1, n=0$) with an anormalously small frequency appears and the
doublets with $m=\pm |m|$ are split.

For consideration of the vortex state it is natural to write the
magnetization, $ \vec M$ by use of angular variables in the polar
coordinate system $M_x=M_s\sin \theta \cos \varphi $, $M_y=M_s\sin
\theta \sin \varphi $, $ M_z=M_s\cos \theta $. For thin enough
dots with $L$ smaller that the vortex core size the magnetization
is uniform along the dot axis $z$, and the vortex state is given
by the ansatz

\begin{equation}
\theta =\theta (r),\varphi =\chi \pm \varphi _0.  \label{ansatz}
\end{equation}
where $r$ and $\chi $ are polar coordinates in the dot plane.

This distribution is typical for magnetic vortices in
two-dimensional easy plane (EP) ferromagnet (FM), and the function
$\theta (r)$ is determined  by numerical solution of an ordinary
differential equation.\cite{Kosevich6} This gives the results that
in the center of the dot $(r=0) \sin \theta (r)=0$ and for $r \gg
l_0$, the value $\theta (r)$ tends to $\pi /2$ exponentially. For
the magnetic dot made of soft magnetic materials the
crystallographic EP anisotropy is negligible, and the
demagnetization field, $\vec H_m$ is dominant. Assuming that the
magnetization along the dot axis $ z $ is uniform, the energy is
as an integral over the dot plane,

\begin{equation}
W=\frac 12L\int d^2x\left[ \left( A/M_s^2\right) \left( \nabla
\vec M\right)^2-\vec M\vec H_m\right]
\label{energy}
\end{equation}

The sources of magnetostatic field $\vec H_m=-\nabla \Phi $ are
both volume and surface "magnetic charges" arising from ${\text{
div}} \vec M$, and from the discontinuity of the normal component
of $\vec M$ on the surface. For determination of the demagnetizing
field the vortex distribution (\ref{ansatz}) is very simple
because ${\text{ div}} \vec M=0$ and $\vec M\vec{\hat{r }}|_{r=R}$
for $\varphi _0=\pi /2$, thereby giving neither dot edge nor
volume contributions and the sole source of field, for the ground
state is $ M_z$. For a thin enough dot this gives the contribution
to the energy $-\vec M\vec H_m/2=2\pi M_z^2$ which is an effective
EP anisotropy forming the out-of-plane vortex structure. The
magnetostatic potential from the edge and volume are determined by

\begin{equation}
\Phi _v=\int \frac{{\text{div}}\vec M}{\left| \vec r-\vec r\
^{\prime }\right| } d^3x,\Phi _{edge}=\int\limits_{r=R}\frac{(\vec
M\vec{\hat{r}}\ ) }{\left| \vec r-\vec r\ ^{\prime }\right|
}Rd\chi dz \label{potential}
\end{equation}
where $( \vec r-\vec r\ ^{\prime }) ^2=r^2+r^{\prime
2}-2rr^{\prime }\cos ( \chi -\chi ^{\prime }) +( z-z^{\prime })
^2$. The magnon modes are investigated using the usual procedure
of expansion in small deviations from the vortex state using
$\theta =\theta _0(r)+\vartheta $ and $\varphi =\chi +\pi /2+\psi
$. It is convenient to define the new small variable, $\mu =-\psi
\sin \theta _0$ so that the corrections to magnetization can be
expressed in the simple form, $ \delta \vec M/M_s=\vartheta \left(
\vec{\hat{\chi }}\cos \theta _0-\vec{\hat z }\sin \theta _0\right)
+ \mu \vec{\hat{r}}$. The linear equations of motion for $\mu ,
\vartheta $ can be written as

\begin{equation}
4\pi M_s\widehat{G} {\vartheta \choose \mu} -\frac 1r {\cos \theta
_0\left( \partial \Phi /\partial \chi \right)  \choose r\left(
\partial \Phi /\partial r\right) } =\frac 1\gamma \frac \partial
{\partial t} {-\mu  \choose \vartheta} \label{G}
\end{equation}
where $\Phi =\Phi _v+\Phi _{edge}$ , from (\ref{potential}), and
$\widehat{G}$ is the same dimensionless operator as for the vortex
in EP magnets.\cite {Ivanov4} Thus, in contrast to the case of EP
magnets with local interactions, \cite{Ivanov4} the equations for
and $\vartheta $ and $ \mu $ become integro-differential.

To proceed with the solution of (\ref{G}) the ansatz $\vartheta
=f_m(r)\cos (\chi m+\omega t)$, $\mu =g_m(r)\sin (\chi m+\omega
t)$ is used. It is easy to see that for this ansatz the values of
$ \nabla \vec M$ and $\vec M\vec{\hat{r}}|_{r=R}$ , as well as the
magnetostatic potentials $\Phi _v$ and $\Phi _{edge}$ , are
proportional to $\sin (\chi m+\omega t)$. Then, one obtains the
set of ordinary integro - differential equation for
functions$f_m(r),g_m(r)$ instead of (\ref{G}) with partial
derivatives.

The edge contribution to $\vec H_m$ can be estimated from the
integral (\ref{potential}), which diverges as $L\rightarrow \ 0$
giving a non-analytic contribution to the system energy for small
$L$. Using asymptotic techniques, the edge contribution to the
magnetostatic field can be written in the form $H_r =-\mu (R,\chi
)M_sF_m(r)$, where $ F_m(r)=(L/R)(r/R)^{|m|-1}$ as $r\rightarrow \
0$, and reaches its maximal value independent of the small
parameter $L/R$, with $F_m(r)\rightarrow $\ $ 2\pi $ , as
$r\rightarrow \ R$ near the edge. The maximal value of the volume
contribution to $H_r$  is small (the order of $L/R$) over all of
the dot. Thus, the main contribution to the magnetostatic energy
is determined by the field $H_r$ in the region near the edge $R-r
\leq  L$. The leading (logarithmic) contribution to the energy in
the parameter $L/R$ for any $m$ is determined from the value of
$g$ on the edge only,

\begin{equation}
W_{edge}=2\pi L^2RI_mM_s^2g^2(R)\ln \left( \frac{4R}L\right)
\label{Wedge}
\end{equation}
where $I_m=2$ for $m=0$ and $I_m=1$ for the other modes.
Therefore, to lowest order in $L/R$ , the contribution from
non-local magnetostatic potentials can be accounted by use of the
effective boundary condition for $ g(r)$,

\begin{equation}
R\frac{dg(r)}{dr}\mid _{r=R}+g(R)\Lambda =0,\Lambda
=\frac{RL}{2\pi l_0^2} \ln \left( \frac{4R}L\right)
\label{Boundary}
\end{equation}

To find the frequencies of magnon modes in the lowest
(logarithmic) order in $L/R$ , one can solve  (\ref{G}) with $\Phi
=0$, find $\mu (r)$ and apply the boundary condition
(\ref{Boundary}). The solution far from the vortex core, $ r \gg
l_0$ is $g_m(r)=J_m(kr)+\sigma _m(kl_0)Y_m(kr)$ , where
$J_m(z),Y_m(z)$ are Bessel and Neumann functions, respectively,
$k$ is a wave number, and $ \sigma _m(kl_0)$ is the so-called
scattering amplitude.\cite{Ivanov4} For the case of interest $kR
\leq  1$ the value of $kl_0 \leq  l_0/R  \ll 1$ , and long-wave
approximation is valid. In this case, the spin wave frequency
$\omega =\omega _0 l_0k$, where $\omega _0 =4\pi \gamma M_s$, and
the asymptotics for $\sigma _m(kl_0)$ found in \cite{Ivanov4} can
be used. For estimates, we will use the values for permalloy,
$A^{1/2}/M_s= 17$ nm, that gives $l_0=4.8$ nm and the
characteristic demagnetizing field $4\pi M_s=1.012$ T, leading to
the following results.

In the first approximation on $kl_0 \ll 1$ the scattering
amplitude is small and $g_m(r)\cong J_m(kr)$. The frequencies of
modes with $ m=|m|$ and $m=-|m|$ are equal and are determined by
$\omega _{n,m}=\varpi _{n,\left| m\right| }=4\pi \gamma M_s\left(
\kappa _{n,m}l_0/R\right) $, where $\kappa _{n,m}$ is the $n$-th
solution of $\kappa J_m^{\prime }(\kappa )+\Lambda J_m(\kappa
)=0$. The value of $\kappa _{n,m}$ falls between the $n$ -th root
of the Bessel function $j_{m,n}$ and the $n$-th root of its
derivative $j_{m,n}^{\prime },\kappa \rightarrow $ $j_{m,n\text{
}}$ at $\Lambda  \gg m$ and $\kappa \rightarrow   j_{m,n\text{
}}^{\prime }$in the opposite case. Taking into account $\sigma
_m$, the value of which depends on the sign of $m$,\cite{Ivanov4}
produces the splitting of these doublets for the modes with $m\neq
0$. The scattering amplitude is maximal (linear in $k$) for the
modes with $m\pm 1$, with $\sigma _m=\pi kl_0m/4\left| m\right|$,
\cite{Ivanov4} and the splitting of this doublet is the order of
the next power on the small parameter $l_0/R$, $\omega
_{m=-1}-\omega _{m=1}\approx 6.16\omega _0(l_0/R)^2$. Thus, for a
typical $R$ like $100$ nm, this splitting is of order $(0.5\div
2)$GHz and is much smaller than the mean frequency, see Fig.1.

Large values of the scattering amplitude leads to the most
interesting feature of the vortex state dot frequencies, namely,
the appearance of a low lying mode with $m=1$ and $n=0$
corresponding to vortex displacement from the center of the dot,
the so-called translational Goldstone mode (TGM), for which
$J_1(kR)\sim (kl_0)Y_m(kR)$ and $kR \ll 1$. \cite{Ivanov4} For
vortex state magnetic dots the frequency of the TGM mode, shown in
Fig.2 , depends non-monotonically on $R$, $\omega _{TGM}=\omega
_0(l_0/R)^2(\Lambda -1)/(\Lambda +1)$. Comparison of Figs. 1 and 2
shows that it has the same order of magnitude as the doublet
splitting for higher modes with $|m|=1$, and it much smaller than
$\varpi _1$ or smaller than frequencies for homogeneously
magnetized dots. The frequency of TGM  found from the effective
equation of motion for the coordinate of the vortex $\vec X$ with
some model assumptions about static vortex energy shows different
dependence on the dot radius,\cite {Usov7, Guslienko8} namely,
$\omega \propto 1/R^3$ in Ref. \onlinecite {Usov7} and $\omega
\propto 1/R$ in Ref. \onlinecite{Guslienko8}. In contrast
with,\cite {Usov7, Guslienko8} we do not need in any model
assumption and our approach is based on the analysis of
Eq.(\ref{G}).

The modes with $m$ = 0 are singlets having cylindrical symmetry,
and their frequencies for $n$ = 0 and $n$ = 1 are shown in Fig. 1.
The lower mode with $m=0,n=0$ can be considered as coupled
oscillations of the vortex core size and $\varphi _0$ defined in
(\ref{ansatz}). It can be observed in resonance experiments with
an ac field perpendicular to the dot plane.

Modes with $|m|>1$ have small values of $\sigma _m\propto
(kl_0)^\eta $, $\eta \geq 4$ for different $m$ with small doublet
splitting, $\left( \omega _{n,-m}-\omega _{n,m}\right) /\varpi
_{n,\left| m\right| }\approx \left( l_0/R\right) ^\eta $, that
cannot be seen on the scale used for Fig. 1. The mean frequencies
$\varpi _{n,\left| m\right| }$ of these modes for the vortex state
dots have the same order of magnitude as for the homogeneous
states modes calculated in.\cite{Guslienko5}

In summary, it is shown that the dynamics of the magnetic dot in
the vortex state differs significantly from the case of uniformly
magnetized dots. The magnon mode corresponding to oscillations of
the vortex position (TGM, $m = 1, n = 0$) have the lowest
frequency. The next mode $(m = 0, n = 0)$ describes the
oscillations of the vortex core size. These two modes can
potentially be detected in resonance experiments, with ac - field
parallel and perpendicular to the dots plane, respectively. Higher
modes with $m \neq 0$ forms doublets with maximal splitting less
that 1 GHz.

\begin{acknowledgements}

We thank A.N. Slavin, D.D. Sheka and N.A. Usov for useful
discussions. This work is supported by NSF grants DMR-9974273,
DMR-9972507, and INTAS Foundation grant No 97-31 311.

\end{acknowledgements}

\begin{figure}
\includegraphics[width=5 in]{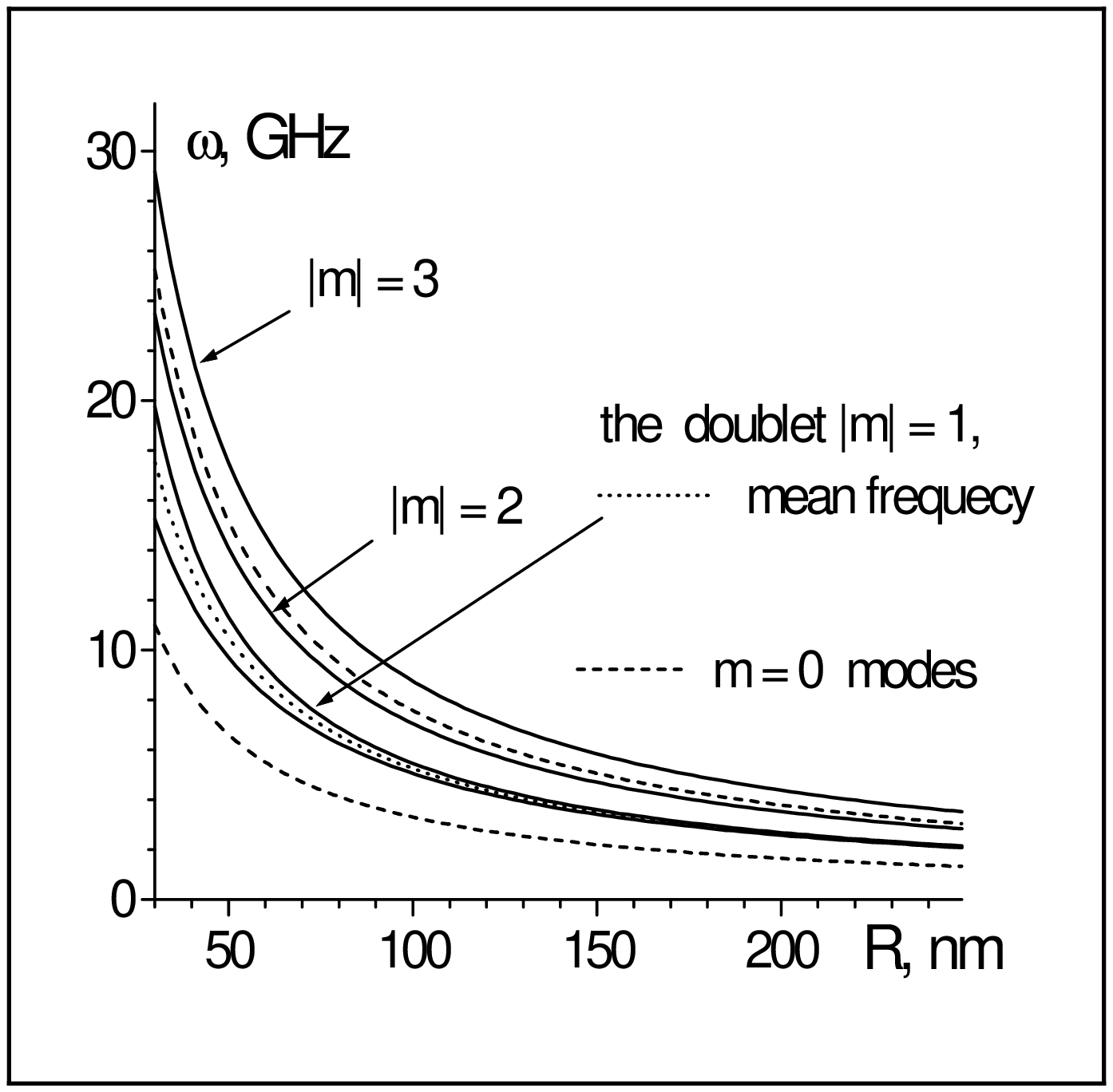}
\caption{ \label{Fig1}   Mode frequencies as a function of dot
radius $R$.}
\end{figure}

\begin{figure}
\includegraphics[width=5 in]{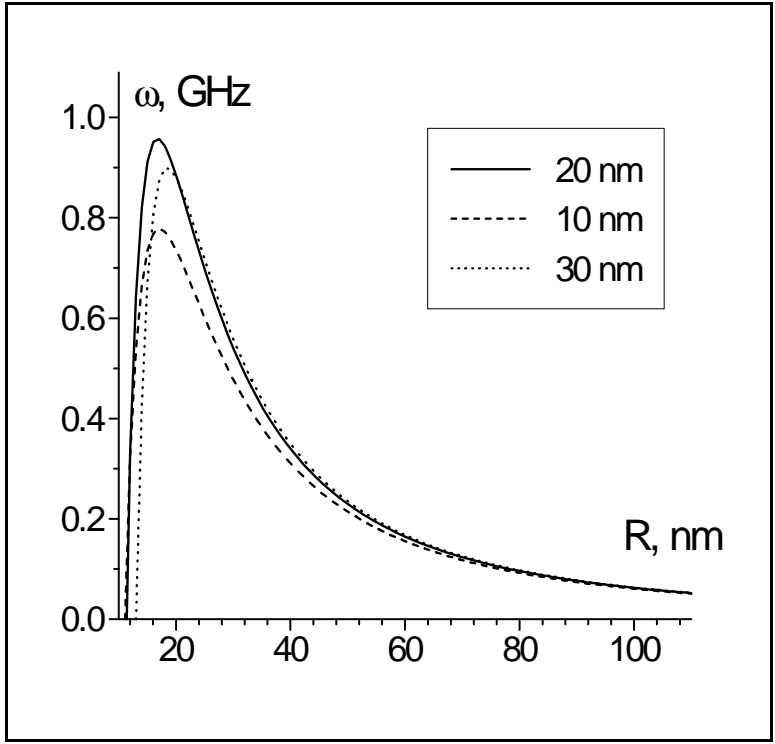}
\caption{ \label{Fig2}   Frequency of the TGM mode versus dot
radius $R$ for different values of dot thickness $L$.}
\end{figure}

\end{document}